\documentclass[aps,
               prl,
               reprint,
               superscriptaddress,
               amsfonts,
               amsmath,
               amssymb,
               floatfix]{revtex4-2}
\RequirePackage[activate={true,nocompatibility},
    final,
    tracking=true,
    factor=1100,
    stretch=10,
    shrink=10]{microtype}
\RequirePackage[binary-units = true,
    range-units = single,
    list-units = single,
    detect-all = true,
    multi-part-units=single]{siunitx}
\DeclareSIUnit\pixel{px}
\RequirePackage{
    graphicx, 
    multirow, 
    mathrsfs,
    longtable, 
    mathtools, 
    etoolbox,
    nicefrac
    }
\RequirePackage[pdfpagemode={UseOutlines},
    bookmarksopen=true,
    bookmarksopenlevel=0,
    hypertexnames=false,
    hidelinks,
    colorlinks=false,% Set to false to disable coloring links
    citecolor=mdtGrey,% The color of citations
    linkcolor=mdtGrey,% The color of references to document elements (sections, figures, etc)
    urlcolor=mdtGrey,% The color of hyperlinks (URLs)
    unicode,
    breaklinks=true]{hyperref}
\RequirePackage{newtxtext,newtxmath}
\RequirePackage[T1]{fontenc} 
\makeatletter
\patchcmd{\section}{\centering}{\raggedright\normalsize}{}{}
\makeatother

\begin{document}

\preprint{v1.0}
\title{Finding the semantic similarity in single-particle diffraction images\\using self-supervised contrastive projection learning}

\author{Julian Zimmermann}
\email{jzimmermann@phys.ethz.ch}
\affiliation{ETH Z\"urich, 8092 Z\"urich, Switzerland}%

\author{Fabien Beguet}
\affiliation{ETH Z\"urich, 8092 Z\"urich, Switzerland}%

\author{Daniel Guthruf}
\affiliation{ETH Z\"urich, 8092 Z\"urich, Switzerland}%

\author{Bruno Langbehn}
\affiliation{Technische Universit\"at Berlin, 10623 Berlin, Germany}%

\author{Daniela Rupp}
\affiliation{ETH Z\"urich, 8092 Z\"urich, Switzerland}%
\affiliation{Max-Born-Institut, 12489 Berlin, Germany}%

\date{\today}

\begin{abstract}
    Single-shot diffraction imaging of isolated nanosized particles has seen remarkable success in recent years, yielding in-situ measurements with ultra-high spatial and temporal resolution. The progress of high-repetition-rate sources for intense X-ray pulses has further enabled recording datasets containing millions of diffraction images, which are needed for structure determination of specimens with greater structural variety and for dynamic experiments. The size of the datasets, however, represents a monumental problem for their analysis. Here, we present an automatized approach for finding semantic similarities in coherent diffraction images without relying on human expert labeling. By introducing the concept of projection learning, we extend self-supervised contrastive learning to the context of coherent diffraction imaging. As a result, we achieve a semantic dimensionality reduction producing meaningful embeddings that align with the physical intuition of an experienced human researcher. The method yields a substantial improvement compared to previous approaches, paving the way toward real-time and large-scale analysis of coherent diffraction experiments at X-ray free-electron lasers.
\end{abstract}
\maketitle

\section{I\lowercase{ntroduction}}\vspace{-1em}
\label{sec:introduction}
\noindent
A guiding principle in fundamental condensed matter research is that for understanding function, we have to study structure \cite{Note1}. Techniques based on lensless diffractive imaging, like X-ray crystallography and coherent diffraction imaging (CDI), are powerful and widely used tools to discover structures up to atomic resolutions \cite{Miao2015}. In the recent past, single-particle coherent diffraction imaging using intense coherent X-ray pulses from free-electron lasers (SP-CDI) \cite{Chapman2010,Miao2015} has revolutionized the field of structural characterization \cite{Seibert2011,Bostedt2010,Loh2012,Xu2014,Gorkhover2012,Gomez2014,Barke2015,Ekeberg2015,Langbehn2018}. SP-CDI is a technique with which in-situ measurements of isolated and non- fixated nano-scaled targets can be acquired. Depending on the experimental scheme, each recorded diffraction image is a complete and self-contained experiment that needs individual analysis \cite{Chapman2010}. However, due to the advent of high repetition-rate sources like the European XFEL \cite{Tschentscher2017} and LCLS-II \cite{Stohr2011}, millions of images are typically recorded during one experimental campaign \cite{Ayyer2021}.
Manual analysis of such amounts of data represents an enormous problem. It may leave researchers unable to analyze significant amounts of their data comparatively as they have to resort to large-scale averaging, which might \emph{wash out} or \emph{conceal} important information, or manually select subsets of the dataset. In this work, we present a novel embedding technique for diffraction images called contrastive projection learning (CPLR) based on contrastive learning (CLR) \cite{Chen2020a,Chen2020b}. CPLR produces a dimensionality-reduced embedding space with which semantic comparisons between diffraction images become possible and, thus, enables human-level comparative analysis on big-data scale datasets.

At the very core of every comparative analysis is an assumption establishing a \emph{similarity measure} between samples. However, current approaches for establishing such a measure for diffraction images cannot compete with the perception of a trained researcher. This perceived similarity, or \emph{semantic similarity} \cite{Note2}, is contextually aware \cite{Note3}, whereas computational methods for diffraction image data currently lack such awareness.

Available strategies for comparative analysis are based on either supervised classifier schemes \cite{Bobkov2015,Zimmermann2019} or unsupervised sorting methods \cite{Yoon2011,Park2013,Andreasson2014,Rose2018,Zhuang2022}. However, all these methods come with significant trade-offs: Supervised algorithms align with human perception and produce high-accuracy results \cite{Zimmermann2019,Ribeiro2016} but are in real-world scenarios unavailable as they require time- and labor-expensive manually fabricated expert labels. Unsupervised routines work in such cases but do not reach comparable accuracy levels and introduce additional restrictions or requirements. For example, traditional cluster techniques produce a lot of unwanted predictions \cite{Yoon2011}, threshold-based approaches act primarily as hit-finder \cite{Park2013,Barty2014}, autocorrelation-based methods only extract particle size information and are computationally costly \cite{Andreasson2014}, auxiliary approaches rely on rarely available additional experimental data \cite{Andreasson2014}, and Fourier-inversion based techniques \cite{Andreasson2014,Zhuang2022,Ayyer2021} are only applicable to SP-CDI data from the \emph{small-angle-scattering} regime, where reconstruction by Fourier inversion is possible \cite{Barke2015,Colombo2022}.

Our CPLR method can potentially improve all strategies mentioned above; It can serve as an improved similarity measure for unsupervised methods, as in the context of regular self-supervised learning \cite{Zhuang2019,Caron2020,VanGansbeke2020}, and can act as a powerful pretraining for subsequent supervised or distillation-based training \cite{Chen2020b}. Furthermore, CPLR directly establishes a way to find \emph{semantically similar} diffraction images in a fully self-supervised fashion. In self-supervised contrastive learning, a supervised task is constructed by artificially creating label information via domain-specific augmentation strategies \cite{Robinson2021,Chen2022,Chen2020a}. In this work, we design a novel augmentation approach for diffraction image data where a deep neural network contrasts images from different coordinate projections.

The quality of the CPLR embedding space is evaluated using a publicly available diffraction image dataset \cite{Note4} from an SP-CDI experiment on superfluid helium nanodroplets, for which semantically sensitive expert labels are available \cite{Langbehn2018,Zimmermann2019}. Using the broadly established linear evaluation protocol \cite{VanDenOord2018,Chen2020a,Chen2020b}, we show that our method outperforms non-contrastive methods by a large margin while improving the contrastive-learning baseline by \SIrange{6}{10}{\percent}.

\begin{figure}[t!]
    \includegraphics[width=\columnwidth]{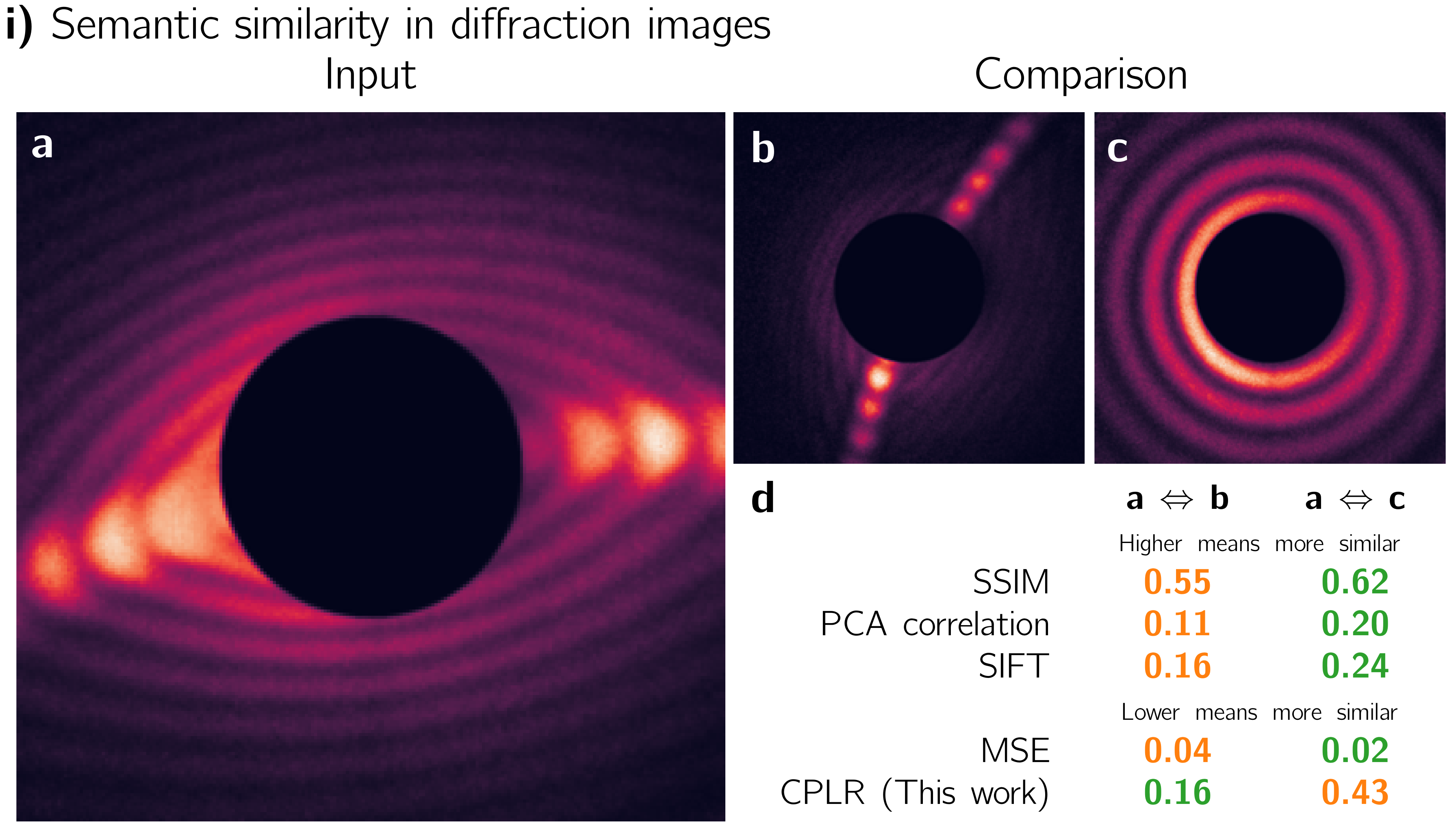}
    \caption{\label{fig:mse_vs_ssim}
    Illustrating semantic similarity.
        {\bf a}, {\bf b} and {\bf c} are diffraction images taken from a publicly available dataset \cite{Note4} from an SP-CDI experiment on superfluid helium nanodroplets \cite{Langbehn2018,Zimmermann2019}. {\bf a} and {\bf b} are semantically similar while {\bf a} and {\bf c} are not.
        {\bf d} shows four widely used similarity measures (see \cite{Note5}), that disagree with the human perception, while only our method (labelled CPLR) agrees with it. Color-coding is that the pair of images that is \emph{more-similar} is green, while the one that is \emph{less-similar} is orange.
    }
\end{figure}%
Figure \ref{fig:mse_vs_ssim} provides a concrete example of three diffraction images from \cite{Langbehn2018}, {\bf a}, {\bf b}, and {\bf c}. Two images ({\bf b} and {\bf c}) are to be compared to {\bf a}. A human immediately identifies the elongated streak-like feature in {\bf a} as the dominant characteristic and can identify {\bf b} as being more similar to {\bf a} than {\bf c} is to {\bf a}. And, indeed, this is correct from a physical perspective. The nanoparticles' structures that produce {\bf a} and {\bf b} are more similar to each other than those that produce {\bf a} and {\bf c} \cite{Langbehn2018}. However, from an algorithmic perspective, this is not a trivial problem. Figure \ref{fig:mse_vs_ssim} {\bf d} shows for {\bf b} and {\bf c} five similarity measures: Four widely used measures for image similarity (see \cite{Note5}) and our method (labeled CPLR). Color-coding is that the image that is \emph{more similar} to a is green, while the \emph{less similar} image is orange. The only measure that agrees with the human perception, and the physics of the problem, is our CPLR method.

Our approach is not limited to data from SP-CDI experiments. Theoretically, CPLR can be applied to diffraction data from all experimental techniques operating in Polar coordinates, including X-ray crystallography and traditional CDI approaches. Ultimately, CPLR provides a path for analyzing the impending amounts of diffraction data where the human perceived similarity is maintained even among millions of diffraction images.

\section{R\lowercase{esults and Discussion}}
\label{sec:results}
\noindent
\textbf{Contrastive learning is about augmentation, not architecture.}
In contrastive learning (CLR), we artificially create label information for supervised learning by designing augmentation pipelines that consider domain \cite{NoteDomainKnowledge} and task \cite{NoteTaskKnowledge} knowledge \cite{Chen2020a}. Therefore, CLR is an instance of self-supervised learning \cite{NoteSelfSupervised}. The fundamental assumption in self-supervised learning is that the input data contain more task-specific information than sparse categorical ground truth data in supervised learning \cite{Liu2021}. Consequently, a careful augmentation design should provide better results on downstream tasks than a supervised learning scenario \cite{Becker1992,Liu2021}. While improvements in \emph{Accuracy} in supervised learning are usually related to architecture modifications, regularization, or loss function, CLR is about domain-specific augmentation strategies above anything else \cite{Chen2020a,Liu2021}. Formally, CLR is a technique to create an embedding space from arbitrary input modalities, enabling comparative analysis. CLR dates back to work done in the nineties \cite{Becker1992} but only recently has seen a renaissance, yielding State-of-the-art results in visual- \cite{VanDenOord2018,Chen2020a,Chen2020b,Tomasev2022}, audio- \cite{AlTahan2020,Wang2021,Saeed2021}, video- \cite{Liu2022,Dave2022,Pan2021}, and text-representation \cite{Gao2021,Rethmeier2021} learning.

\noindent
\textbf{Contrastive baseline and contrastive projection learning.}
In this study, we use the experimental design presented in \cite{Chen2020b}, called \emph{SimCLRv2}, as a baseline to compare our results. A large encoder and a smaller transformation neural network produce the representations in two stages. First, the encoder acts as a feature extractor; then, the transformation network learns an optimized representation that minimizes the CLR loss, termed normalized temperature-scaled cross-entropy loss (NT-Xent) \cite{Sohn2016}. Conceptually, a duplicate is produced for each input image where both images are heavily augmented. All duplicates form so-called \emph{positive pairs} with their originals, while all images with all other images but their duplicates form \emph{negative pairs}. Then, the network learns to discriminate between positive and negative pairs during training.

\begin{figure*}[t!]
    \includegraphics[width=\linewidth]{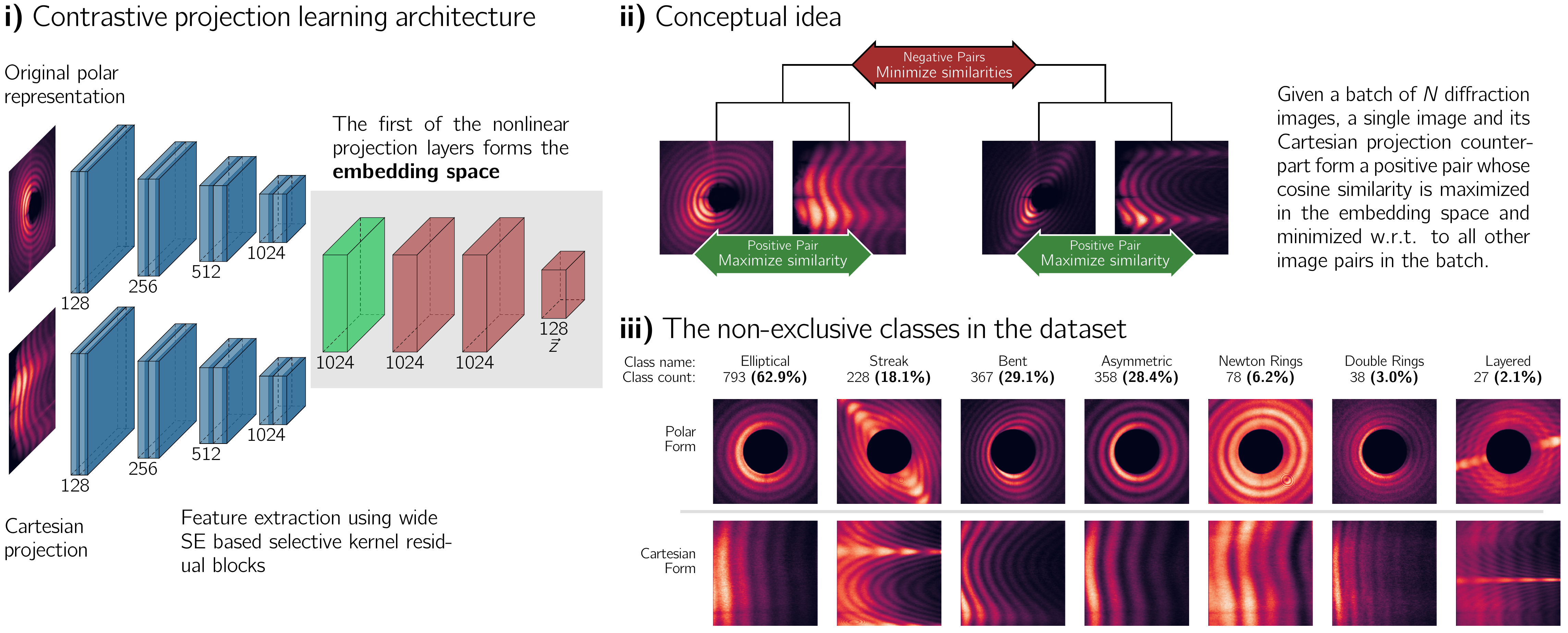}
    \caption{\label{fig:exp_scheme}
        Showing details on the CPLR architecture, the concept, and the classes within the dataset. {\bf i)} Schematic of the used architecture. In this example, a positive pair, consisting of one diffraction image in its natural Polar and its Cartesian form, are passed on to the feature-extractor network, a modified ResNet50-D \cite{He2018}. The adjacent grey box then shows the transformation network, a three layer MLP, where the first layer - in green - forms the embedding space within the here used SimCLRv2 \cite{Chen2020b} framework. {\bf ii)} The conceptual idea in more detail. The goal during training is to maximize similarity (minimize distance in embedding space) for positive pairs and minimize similarity (maximize distance in embedding space) for negative pairs. {\bf iii)} The seven possible, non-exclusive classes of the here used multiclass and multilabel dataset. The top row shows the diffraction image in Polar coordinates and the bottom row the Cartesian projection. In addition, the absolute and relative class counts are given alongside the class names. Several different features can appear in a single image which then belongs to several classes simultaneously. For example, a \emph{Streak} feature is also present in the example for the \emph{Layered} class.
    }
\end{figure*}%
In contrastive projection learning (CPLR), we produce the positive pairs not from the same image, as in \cite{Chen2020a,Chen2020b}, but project the diffraction images, which are naturally recorded in Polar coordinates, to Cartesian coordinates. Figure \ref{fig:exp_scheme} {\bf i)} and {\bf ii)} provide a schematic overview of the network design and the conceptual idea. Using coordinate projections as an augmentation strategy implicitly penalizes that trivial Polar symmetries are learned and explicitly enforces that learned representations are invariant under rotational and translational changes. A simple example can be constructed with the help of figure \ref{fig:exp_scheme} {\bf iii)}. There, example images for every class in the dataset are shown. While figure \ref{fig:exp_scheme} {\bf iii)} is fully explained in the following subsection, here, we concentrate on the Polar and Cartesian projection of the \emph{Elliptical} class - the first column. The Polar form shows the characteristic \emph{Airy} rings typical for single-laser-shot and single-particle imaging data \cite{Langbehn2018}. Usually, learning rotational invariance in arbitrary image data is achieved via random rotational transformations during the augmentation stage of training a network. However, such a transformation would yield no, or very little, change with such diffraction images due to the high degree of rotational symmetry. Therefore, we encourage the network to decouple the learned representation from rotational symmetries by correlating the Polar form with the Cartesian form. In addition, after the coordinate transformation, we leverage a stochastic augmentation pipeline \cite{Chen2020a,Chen2020b}. More details can be found in the \emph{Methods} section \emph{Augmentation strategy}.

\noindent
\textbf{The dataset.}
Helium nanodroplets were imaged at XUV photon energies in a single-shot single-particle experiment \cite{Langbehn2018} at the FERMI free-electron laser \cite{Allaria2012}. The scattering images were recorded using a non-linear MCP-type detector \cite{Bostedt2012} in a so-called wide-angle setting where each diffraction image contains 3D-structural information and cannot be reconstructed via Fourier inversion \cite{Barke2015,Rupp2017,Langbehn2018}. The publicly available and hand-curated dataset \cite{Note4} contains \num{7264} diffraction images with semantically sensitive labels and was previously used in a supervised classification task \cite{Zimmermann2019}. We discarded \num{6000} diffraction images as they either exhibited strictly round or no Airy patterns at all. The round Airy patterns are by far the most common class, and we removed them to create a more balanced dataset since they can be reliably sorted using radial slices and a classical peak-finder \cite{Note6}.

The provided expert labels can be used for multiclass and multilabel analysis, meaning every diffraction image has binary label information for multiple classes that are often mutually non-exclusive. Figure \ref{fig:exp_scheme} {\bf iii)} shows this dataset's seven possible classes and their absolute and relative occurrence. For every class, one example is given in Polar and Cartesian form. To illustrate the multilabel property: The characteristic streak-like feature that defines the \emph{Streak} class can also be found, for example, in the image for the \emph{Layered} class. For this reason, the given percentages do not add up to \SI{100}{\percent}, as multiple images belong to multiple classes and most classes are heavily under-represented. This pronounced multilabel imbalance is typical for diffraction image datasets \cite{Langbehn2018,Gomez2014}.

\noindent
\textbf{Training and evaluation.}
As in \cite{Chen2020b}, we use a $\num{2}x$-wide \cite{Zagoruyko2016}, selective-kernel \cite{Li2019} ResNet50-D \cite{He2018} network with squeeze-excitation blocks \cite{Hu2017} as feature extractor, and a three-layer multi-layer- perceptron (MLP) as transformation network. We train for \num{1000} epochs, with a batch size of \num{628}, using the LARS optimizer \cite{You2017} with a cosine schedule \cite{Loshchilov2017}, ten warmup epochs \cite{Goyal2017}, and optimizing the NT-Xent loss \cite{Sohn2016}. Training takes half an hour on four NVIDIA \num{3090} GPUs. The code, pretrained models, and training results are openly available at \cite{Note10}.

The quality of the learned embedding space is evaluated using the so-called linear evaluation protocol \cite{VanDenOord2018,Chen2020a,Chen2020b}, which is carried out as follows: After self-supervised training, we freeze the feature extractor network and use the first layer of the transformation network (this is indicated by the green layer in \ref{fig:exp_scheme} {\bf i)}) as embedding space. Then, for every ground truth class in the dataset, we train a linear classifier on top of the learned representations and calculate the \emph{Precision} and \emph{Recall} score. Both metrics are obtained for every class via \num{5}-fold stratified cross-validation to account for statistical fluctuations from sampling and the dataset's class imbalance. It has turned out to be important to use \emph{Precision} and \emph{Recall} as most classes are very rare, with three out of seven classes appearing in under \SI{7}{\percent} of all images. In these cases, \emph{Accuracy} would produce very high scores for classifiers predicting every image as not being part of any class.
Moreover, we calculate an additional metric called \emph{Overlap} in order to compare our method with metrics operating on raw images, such as the structural similarity (SSIM) index \cite{Wang2004}, the complex wavelet SSIM (CWSSIM) index \cite{Wang2005}, and the keypoint-based scale-invariant feature transform (SIFT) distance \cite{Lowe1999}. \emph{Overlap} is the global average of the normalized dot-product between the ground truth of every image and its \num{13} closest images according to the pairwise-calculated distances. \num{13} was chosen as it corresponds to \SI{1}{\percent} of all images in the dataset, which sets the \emph{Overlap} score to be a local-neighborhood evaluation. Consequently, an \emph{Overlap} score of \num{0.5} corresponds to: \emph{On average, the \num{13} most similar images shared \SI{50}{\percent} of the original image's labels}.

We compare our method to the SimCLRv2 baseline and other approaches that have been used in the past with diffraction images in the following sections. More details on training and evaluation can be found in the \emph{Methods} section \emph{Training and linear evaluation strategy}.

% \\~\\
\noindent
\textbf{The embedding space is linearly separable into semantic features.}
\begin{table}[t!]
    \caption{\label{tbl:macro_avg_clr}
        Macroaverage \cite{NoteMacroMicroAveraging} results on the helium nanodroplets dataset.
        \emph{Contrastive-based} are the results for the CLR baseline and the CPLR method, where the arrows indicate with which projections the contrastive task was constructed. \emph{Continuous latent variables} or \emph{variational Bayesian} methods list techniques that have been used with diffraction data in the past. \emph{Random baseline} gives the result for an artificial embedding space built from uniform noise, this is the lowest possible score. \emph{Direct measures} are applied directly on the images, and cannot be evaluated using the linear evaluation protocol. All methods except for the \emph{Direct measures} are trained / ran for five times where each time the evaluation scores where obtained via \num{5}-fold stratified cross validation. The standard deviation of all methods is equal or below \num{0.01}. The best result for each score is given in bold letters.
    }
    \begin{tabular}{ccccc}
        {}                                      & \multicolumn{2}{c}{{\bf Linear evaluation}} & \multicolumn{2}{c}{{\bf Local similarity}}                                          \\
        {\bf Method}                            & Precision                                   & Recall                                     & Measure             & Overlap          \\[-.25cm]
        \multicolumn{5}{c}{~}                                                                                                                                                       \\
        \multicolumn{5}{c}{{\bf\footnotesize Contrastive-based}}                                                                                                                    \\[-.25cm]
        \multicolumn{5}{c}{~}                                                                                                                                                       \\
        \multicolumn{5}{c}{{\scriptsize CLR}}                                                                                                                                       \\[-.25cm]
        \multicolumn{5}{c}{~}                                                                                                                                                       \\
        Polar$\,\Leftrightarrow\,$Polar         & \num{0.49}                                  & \num{0.50}                                 & Cosine              & \num{0.49}       \\
        Cartesian$\,\Leftrightarrow\,$Cartesian & \num{0.49}                                  & \num{0.48}                                 & Cosine              & \num{0.43}       \\[-.25cm]
        \multicolumn{5}{c}{~}                                                                                                                                                       \\
        \multicolumn{5}{c}{{\scriptsize CPLR}}                                                                                                                                      \\[-.25cm]
        \multicolumn{5}{c}{~}                                                                                                                                                       \\
        Cartesian$\,\Leftrightarrow\,$Polar     & \num{0.51}                                  & \num{0.54}                                & Cosine              & {\bf \num{0.52}} \\
        Polar$\,\Leftrightarrow\,$Cartesian     & {\bf \num{0.52}}                            & {\bf \num{0.55}}                           & Cosine              & {\bf \num{0.52}} \\
        \multicolumn{5}{c}{~}                                                                                                                                                       \\
        \multicolumn{5}{c}{{\bf\footnotesize Continuous latent variables methods}}                                                                                                  \\[-.25cm]
        \multicolumn{5}{c}{~}                                                                                                                                                       \\
        Factor Analysis                         & \num{0.28}                                  & \num{0.35}                                 & Euclidean           & \num{0.28}       \\
        \multirow{2}{*}{PCA}                    & \multirow{2}{*}{\num{0.29}}                 & \multirow{2}{*}{\num{0.35}}                & Euclidean           & \num{0.40}       \\
                                                & ~                                           & ~                                          & Correlation         & \num{0.43}       \\
        \multirow{2}{*}{Kernel PCA}             & \multirow{2}{*}{\num{0.28}}                 & \multirow{2}{*}{\num{0.35} }               & Euclidean           & \num{0.43}       \\
                                                & ~                                           & ~                                          & Correlation         & \num{0.40}       \\
        UMAP \cite{McInnes2018}                 & \num{0.30}                                  & \num{0.31}                                 & Euclidean           & \num{0.41}       \\%[-.25cm]
        \multicolumn{5}{c}{~}                                                                                                                                                       \\
        \multicolumn{5}{c}{{\bf\footnotesize Variational Bayesian methods}}                                                                                                         \\[-.25cm]
        \multicolumn{5}{c}{~}                                                                                                                                                       \\
        VAE \cite{Kingma2013,Burgess2018}       & \num{0.34}                                  & \num{0.35}                                 & Wasserstein $W_{2}$ & \num{0.42}       \\
        \multicolumn{5}{c}{~}                                                                                                                                                       \\
        \multicolumn{5}{c}{{\bf\footnotesize  Random baseline}}                                                                                                                     \\[-.25cm]
        \multicolumn{5}{c}{~}                                                                                                                                                       \\
        Uniform Noise                           & \num{0.23}                                  & \num{0.27}                                 & Euclidean           & \num{0.28}       \\
        \multicolumn{5}{c}{~}                                                                                                                                                       \\
        \multicolumn{5}{c}{{\bf\footnotesize Direct measures}}                                                                                                                      \\[-.25cm]
        \multicolumn{5}{c}{~}                                                                                                                                                       \\
        SSIM \cite{Wang2004}                    & N/A                                         & N/A                                        & Custom              & \num{0.37}       \\
        CWSSIM \cite{Wang2005}                  & N/A                                         & N/A                                        & Custom              & \num{0.36}       \\
        SIFT \cite{Lowe1999}                    & N/A                                         & N/A                                        & Euclidean           & \num{0.32}       \\
    \end{tabular}
\end{table}%
The evaluation scores after training are provided in table \ref{tbl:macro_avg_clr}. The second and third columns show the \emph{Precision} and \emph{Recall} score of the linear evaluation protocol, and the last two columns show the used metric for calculating the pairwise distances to calculate the \emph{Overlap} score, which is given in the last column. \emph{Contrastive-based} shows the results for the CLR baseline and our CPLR method. The arrows indicate the coordinate projections used, where the first term is used for inference and the second is used for constructing the contrastive task. Consequently, \emph{Polar$\,\Leftrightarrow\,$Polar} and \emph{Cartesian$\,\Leftrightarrow\,$Cartesian} are cases of the standard CLR framework with either purely unmodified (\emph{Polar}) or purely projected (\emph{Cartesian}) diffraction images. \emph{Cartesian$\,\Leftrightarrow\,$Polar} and \emph{Polar$\,\Leftrightarrow\,$Cartesian} are cases of our CPLR method, where the difference between the two is that we changed the input for inference at evaluation time to either the \emph{Cartesian} or the \emph{Polar} form.
\emph{Continuous latent variables} and \emph{variational Bayesian} methods are techniques used previously in the context of diffraction images. More details are given in the \emph{Methods} section \emph{Non-contrastive-based methods}. \emph{Random baseline} gives the result for an artificial embedding space built from uniform noise; this is the lowest possible score. This baseline is equivalent to a \emph{Random guesser} with no learned information about the dataset. \emph{Direct measures} are methods applied directly to the images, which cannot be evaluated using the linear evaluation protocol.

All methods except for the \emph{Direct measures} are trained / run five times using five different integer \emph{random\_state} keys where the evaluation scores were each time obtained via \num{5}-fold stratified cross-validation. The macro average \cite{NoteMacroMicroAveraging} over all classes for all train and cross-validation runs is given in the table. The standard deviation of all methods is equal to or below \num{0.01}.

Relative to the CLR baseline, the CPLR method yields significant improvements on all metrics improving \emph{Precision} by \SI{6}{\percent} and \emph{Recall} by \SI{10}{\percent}. Relative to the best non-CLR methods (VAE for \emph{Precision} and all but UMAP for \emph{Recall}), the CPLR method improves \emph{Precision} by \SI{35}{\percent} and \emph{Recall} by \SI{36}{\percent}. In addition, the \emph{Overlap} score is relatively improved by about \SI{6}{\percent} compared to the CLR baseline and \SI{17}{\percent} compared to the two best PCA-based approaches. CPLR is the only method that achieves \emph{Precision} and \emph{Overlap} scores above \num{0.50}.

% \\~\\
\noindent
\textbf{CPLR is more robust with fewer samples.}
The general idea of using the linear evaluation protocol for evaluation is to look for linearly separable regions in the embedding space. Therefore, this method only applies to \emph{one-hot} \cite{Note7} ground truth data, meaning \emph{multiclass but single label}. However, the helium nanodroplets dataset has multiclass and multilabel \cite{Note8} ground truth data, where each image has multiple associated labels. Moreover, as typical in datasets on helium nanodroplets, the dataset is heavily unbalanced, where simpler shapes, like \emph{Elliptical}, dominate other classes \cite{Gomez2014}.
\begin{table*}[t!]
    \caption{\label{tbl:micro_avg_clr}
        Results for every class in the helium nanodroplets dataset. Compared to table \ref{tbl:macro_avg_clr}, we only show the four best-performing methods, namely our CPLR (\emph{Polar$\,\Leftrightarrow\,$Cartesian}) method along with the CLR (\emph{Polar$\,\Leftrightarrow\,$Polar}) baseline, and the VAE, and Kernel PCA (Using the euclidean metric) approaches. All methods have been trained / run for five times where each time the evaluation scores where obtained via \num{5}-fold stratified cross validation. The standard deviation of all methods and for all classes is equal or below \num{0.01}. The best result for each score is highlighted in bold letters.
    }
    \begin{tabular}{rccccc|ccc|ccc|ccc}
        {}                          & {}                 & {}                  & \multicolumn{3}{c}{CPLR}                               & \multicolumn{3}{c}{CLR}          & \multicolumn{3}{c}{VAE}                                & \multicolumn{3}{c}{Kernel PCA}                                                                                                                                                                                                                                                                                                    \\[-0.25cm]
        \multicolumn{15}{c}{~}                                                                                                                                                                                                                                                                                                                                                                                                                                                                                                                                          \\
        {}                          & {}                 & {}                  & \multicolumn{2}{c}{{\bf {\footnotesize Linear eval.}}} & {\bf {\footnotesize Similarity}} & \multicolumn{2}{c}{{\bf {\footnotesize Linear eval.}}} & {\bf {\footnotesize Similarity}} & \multicolumn{2}{c}{{\bf {\footnotesize Linear eval.}}} & {\bf {\footnotesize Similarity}} & \multicolumn{2}{c}{{\bf {\footnotesize Linear eval.}}} & {\bf {\footnotesize Similarity}}                                                                                                          \\
        {\bf {\footnotesize Class}} & $\mathbf{n_{abs}}$ & $\mathbf{n_{rel}}$  & {\footnotesize Precision}                              & {\footnotesize Recall}           & {\footnotesize Overlap}                                & {\footnotesize Precision}        & {\footnotesize Recall}                                 & {\footnotesize Overlap}          & {\footnotesize Precision}                              & {\footnotesize Recall}           & {\footnotesize Overlap} & {\footnotesize Precision} & {\footnotesize Recall} & {\footnotesize Overlap} \\
        Elliptical                  & 793                & \SI{62.9}{\percent} & {\bf 0.79}                                             & {\bf 0.77}                       & 0.65                                                   & 0.78                             & {\bf 0.77}                                             & 0.65                             & 0.71                                                   & 0.71                             & {\bf 0.66}              & 0.68                      & 0.64                   & 0.61                    \\
        Streak                      & 228                & \SI{18.1}{\percent} & {\bf 0.91}                                             & {\bf 0.90}                       & 0.66                                                   & 0.89                            & 0.86                                                   & {\bf 0.67}                       & 0.83                                                   & 0.72                             & 0.61                    & 0.46                      & 0.47                   & 0.57                    \\
        Bent                        & 367                & \SI{29.1}{\percent} & {\bf 0.51}                                             & {\bf 0.53}                       & 0.48                                                   & {\bf 0.51}                       & 0.51                                                   & {\bf 0.49}                       & 0.40                                                   & 0.50                             & 0.44                    & 0.35                      & 0.41                   & 0.41                    \\
        Asymmetric                  & 358                & \SI{28.4}{\percent} & {\bf 0.42}                                             & 0.38                             & 0.55                                                   & 0.38                             & {\bf 0.40}                                             & 0.55                             & 0.33                                                   & 0.33                             & {\bf 0.56}              & 0.29                      & 0.37                   & 0.52                    \\
        Newton Rings                & 78                 & \SI{6.2}{\percent}  & {\bf 0.32}                                             & {\bf 0.34}                       & {\bf 0.51}                                             & 0.24                             & 0.25                                                   & 0.48                             & 0.06                                                   & 0.04                             & 0.48                    & 0.09                      & 0.21                   & 0.44                    \\
        Double Rings                & 38                 & \SI{3.0}{\percent}  & {\bf 0.34}                                             & {\bf 0.40}                       & {\bf 0.45}                                             & {\bf 0.34}                       & {\bf 0.40}                                             & 0.43                             & 0.05                                                   & 0.04                             & 0.43                    & 0.06                      & 0.22                   & 0.41                    \\
        Layered                     & 27                 & \SI{2.1}{\percent}  & {\bf 0.37}                                             & {\bf 0.37}                       & {\bf 0.33}                                             & 0.31                             & 0.32                                                   & 0.30                             & 0.02                                                   & 0.09                             & 0.26                    & 0.08                      & 0.25                   & 0.29                    \\
    \end{tabular}
\end{table*}%
It is, therefore, instructive to look at the individual averages for every class, which are given in table \ref{tbl:micro_avg_clr}. The CPLR method performs significantly better than the CLR baseline and non-contrastive methods in linear evaluation. The most significant improvement is with rarely occurring classes that appear only in $\leq\SI{7}{\percent}$ of all images. VAE and PCA-based techniques fail entirely to place these diffraction images in a linearly separable region of the embedding space, resulting in poor \emph{Precision} and \emph{Recall} scores. However, the CLR baseline also yields limited success in the case of radial symmetry-breaking features like the \emph{Newton Rings} and \emph{Layered} class. There, the diffraction images contain features that either break radial symmetry (\emph{Layered}) or introduce a second radially symmetric feature (\emph{Newton Rings}), cf. figure \ref{fig:exp_scheme} {\bf iii}), which in combination with a low class-count brings the CLR method to its limits. The symmetry-breaking projection of the CPLR method helps in those cases and yields better results when fewer images are available.

\begin{figure}[t!]
    \includegraphics[width=\columnwidth]{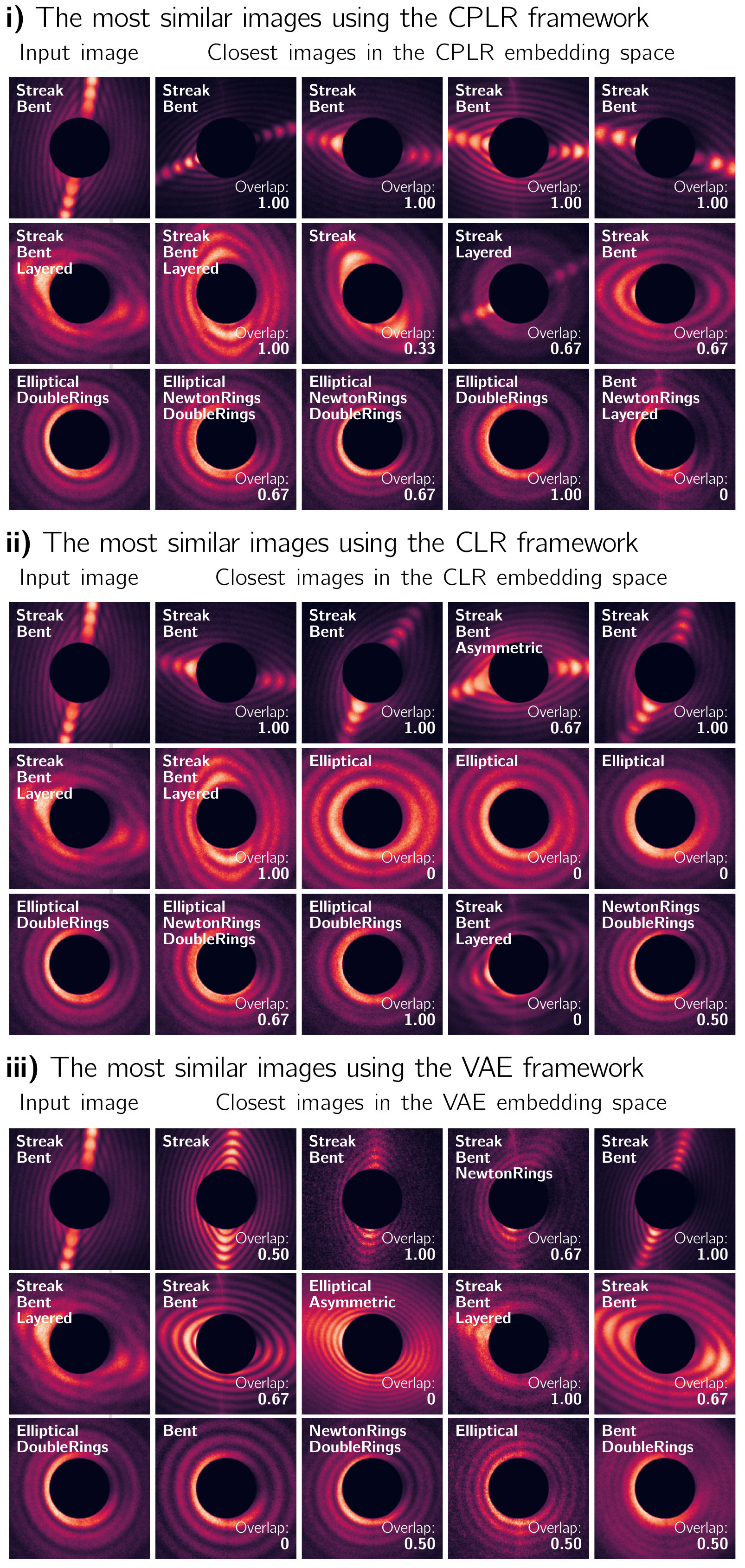}
    \caption{\label{fig:qualitative_results}
        Qualitative results on the helium nanodroplets dataset. In {\bf i)} to {\bf iii)} the column \emph{Input image} shows three randomly chosen images - from a pre-defined class combination. Next to the input images are the four closest diffraction images according to calculated pairwise distances for the CPLR, CLR, and VAE method using the metric given in table \ref{tbl:macro_avg_clr}. Every image is augmented in the top left corner by the class names given in the ground truth data and - for all images but the input image - the \emph{Overlap} score with its corresponding input image.
    }
\end{figure}%
A qualitative comparison of the CPLR, CLR, and VAE results is given in figure \ref{fig:qualitative_results} {\bf i)} to {\bf iii)}. In each plot, the column \emph{Input image} shows the same three diffraction images, randomly chosen out of the three classes \emph{Streak}, \emph{Layered}, and \emph{Double Rings}. Next to each input image are the four diffraction images belonging to the four closest embeddings in the embedding space. Additionally, every image is augmented in the top left corner by the class names given in the ground truth data and, for all images but the input image, by the \emph{Overlap} score with its corresponding input image.

The images for the \emph{Streak} / \emph{Bent} class combination in the first row show strong \emph{Overlap} scores for all three frameworks (\num{1.00} for CPLR, \num{0.92} for CLR, and \num{0.79} for VAE). However, only the contrastive-based methods placed those embeddings of images next to each other where the characteristic \emph{streak} feature is orientated and elongated differently than in the input image. We consider this a strength highlighting that both contrastive methods focus more on the semantics within a diffraction image than the pixel-wise similarity. This can also be seen in the \emph{Streak} / \emph{Bent} / \emph{Layered} class combination in the second row of figure \ref{fig:qualitative_results}, where the direction of the characteristic \emph{Streak} and \emph{Bent} features vary substantially in size and orientation for both CLR-based methods but are identically aligned within the nearest neighbors of the VAE framework.

As already discussed above, our CPLR method outperforms the baseline CLR methods, especially in scenarios with low sample counts and symmetry-breaking features. This behavior is best seen in the second row for the \emph{Streak} / \emph{Bent} / \emph{Layered} class combination. Only \num{27} images in the dataset have a \emph{Layered} label assigned to them, and the CLR method fails to learn this characteristic having an average \emph{Overlap} score of \num{0.25} for this example and only placing one additional image with a \emph{Layered} label near the input image in the embedding space. On the other hand, the CPLR method performs significantly better, with an average \emph{Overlap} score of \num{0.67} and placing two images with a \emph{Layered} label, two with a \emph{Streak}, and one with a \emph{Bent} label next to the input image.

This observation also holds for the third example with the \emph{Elliptical} / \emph{Double Rings} example, where the CPLR method reaches an average \emph{Overlap} score of \num{0.58}, compared to \num{0.54} and \num{0.38} for the CLR and the VAE method, respectively.

These qualitative observations, along with the quantitative results presented above, show conclusively that the CPLR method is introducing significant improvements compared to previous methods for finding the semantic similarity in diffraction images and the baseline CLR method.

\section{S\lowercase{ummary and outlook}}\vspace{-1em}
\noindent
We have introduced a novel method for finding the semantic similarities in diffraction images without relying on expert labeling. Based on contrastive learning (CLR), we introduced contrastive projection learning (CPLR), where the contrastive learning task is constructed from coordinate- projections of an input diffraction image and not from the same image as in CLR. This relatively easy alternation of the learning scenario substantially improves the quality of the learned embedding space on all metrics and scores. CPLR, therefore, provides a much-needed pathway for the upcoming big-data challenges within the coherent diffraction imaging (CDI) community since similarity calculations are at the core of almost every segmentation, classification, and clustering algorithm. Consequently, CPLR can be implemented as a stand-in-replacement for other similarity metrics in all so-far published classification and clustering approaches for diffraction images, potentially improving a wide range of long-established working routines in research groups.

In addition, our method can, theoretically, also be applied to all data that inherit Polar symmetry, such as in X-ray crystallography.

Our results have the potential to enable multiple future possibilities. For example, currently, \num{3}D reconstruction via CDI methods can either be done in the small-angle regime, where reconstruction by Fourier inversion is possible \cite{Barke2015,Colombo2022}, using the \emph{Expand-Maximize-Compress} (EMC) algorithm \cite{Loh2009}, or in the wide-angle regime, via a recursive forward-fitting \emph{Multi-Slice-Fourier-Transform} (MSFT) approach \cite{Barke2015,Rupp2017,Colombo2022}. In both cases, the similarity between diffraction images needs to be calculated. As of today, the EMC method can be applied to datasets on the order of millions of images \cite{Zhuang2022,Ayyer2021}. However, the similarity calculation is currently done using the cross-correlation between radial intensity profile lines of the diffraction images at different angles \cite{Zhuang2022,Ayyer2021}, which is computationally costly, and, as the authors in \cite{Zhuang2022} pointed out, may not be sufficient for more complex patterns. As with the MSFT method, similarity calculations are currently done using either the MSE or even manual estimation by researchers \cite{Barke2015,Langbehn2018,Colombo2022,Colombo2022b}.

Ultimately, CPLR can provide a path to apply the EMC algorithm on more complex datasets, get better results on simpler datasets, and replace the MSE metric in MSFT-based approaches.

Furthermore, it enables quick and reliable statistical reasoning on the variability and occurrence of features within diffraction image datasets, as was done in \cite{Langbehn2018}, for example.

Finally, recent research on contrastive methods in computer vision \cite{Grill2020,Tomasev2022} promises accuracies comparable to supervised methods or surpassing them and can be easily implemented into our framework.

Therefore, this manuscript stands as a stepping stone for adapting self-supervised learning to the domain of diffraction imaging.

All code for the discussed experiments, pretrained models, and extracted embedding spaces are available at our ETH Gitlab repository \cite{Note10}.

\footnotesize
\setlength{\parindent}{0pt}
\section{M\lowercase{ethods}}\vspace{-1em}
\label{sec:methods}
\textbf{Augmentation strategy.}
A well-defined augmentation strategy is critical in contrastive learning \cite{Chen2020a}. As pointed out by \cite{Chen2020a,Chen2020b}, the essential parts of constructing this strategy are random cropping and random color distortion transformations. The latter is targeted towards histogram and color-channel correlation-based overfitting of the network. Since diffraction data is monochrome, we replace the channel-independent RGB distortion with a single-channel jitter distortion. Furthermore, as in \cite{Chen2020a,Chen2020b}, we use a probabilistic augmentation strategy that includes \emph{Flip}, \emph{Rotation}, \emph{Crop \& Resize}, \emph{Jitter}, \emph{Fill}, and \emph{Translation} transformations on all input patches. However, our \emph{Crop \& Resize} routine is not changing the aspect ratio, as is usually done in other contrastive learning augmentation pipelines. Changing the aspect ratio would break the correlation between the \emph{Polar} and \emph{Cartesian} projections. Every transformation has a fixed probability of \SI{50}{\percent} for being applied at every invocation. We implemented the entire pipeline using TensorFlow augmentation layers placed on the GPU itself. Code is available in the official repository \cite{Note10}.

% \\~\\
\textbf{Training and linear evaluation strategy.}
The NT-Xent loss that is minimized during training is given by:
\begin{align}
    \nonumber
    \mathbf{l}_{i,j} = -\log\frac{\exp\left(\text{sim}\left(\mathbf{z}_{i}, \mathbf{z}_{j}\right)/\tau\right)}{\sum^{2N}_{k=1}\mathbf{1}_{[k\neq{i}]}\exp\left(\text{sim}\left(\mathbf{z}_{i}, \mathbf{z}_{k}\right)/\tau\right)},
\end{align}
where $\text{sim}\left(\mathbf{u}, \mathbf{v}\right) = \mathbf{u}^{T}\mathbf{v}/\left(||\mathbf{u}||\,||\mathbf{v}||\right)$ denotes the cosine similarity between two vectors $\mathbf{u}$ and $\mathbf{v}$,  $\mathbf{1}_{[k\neq{i}]} \in$ \{$0, 1$\} is an indicator function evaluating to $1$ if, and only if, $k\neq{i}$, and $\tau$ denotes a temperature parameter. We performed extensive hyper-parameter optimization to obtain the best possible values for the temperature parameter $\tau$, which are \num{0.200} for Polar$\,\Leftrightarrow\,$Polar, \num{0.200} for Cartesian$\,\Leftrightarrow\,$Cartesian, \num{0.075} for Cartesian$\,\Leftrightarrow\,$Polar, \num{0.100} for Polar$\,\Leftrightarrow\,$Cartesian. The results of this hyper-search, as well as scripts to re-run it, can be found in the official repository \cite{Note10}.

The linear classifier we used for linear evaluation was a single-layer perceptron with an \emph{inverse-scaling} learning rate schedule and a l\num{2} penalty of \num{0.0001}. We used the implementation provided by the \emph{sklearn} Python package. Code is available in the official repository \cite{Note10}.

% \\~\\
\textbf{Non-contrastive-based methods.}
Listed in table \ref{tbl:macro_avg_clr} are the \emph{Factor-Analysis} (FA), the \emph{Principal-Component-Analysis} (PCA), the Kernel-PCA, the \emph{Uniform Manifold Approximation \& Projection} (UMAP) \cite{McInnes2018}, and the \emph{Variational Autoencoder} (VAE) \cite{Kingma2013,Burgess2018} methods. All of these have been used with various forms of spectrographic image data. FA- and PCA-based methods are parameter-free dimensionality reduction techniques that are regularly used within all scientific disciplines; while FA considers the dataset's variance, PCA considers the covariance of the data. FA and PCA-based methods have been used with powder diffraction data \cite{Westphal2015,Chernyshov2020} and X-ray diffraction phase analysis \cite{Camara2014} and as a dimensionality reduction for subsequent classification \cite{Matos2007} and clustering \cite{Yoon2011}.

A VAE is a generative \emph{variational Bayesian} model where the input information is encoded to a low dimensional representation via an encoder function and then recreated by a decoder function. The loss function, called the \emph{Evidence lower bound}, is a lower bound on the marginal likelihood \cite{Kingma2013}. VAEs have been used with diffraction images in various tasks, such as anomaly-detection \cite{Banko2021}, dimensionality reduction \cite{RuizVargas2018}, phase reconstruction \cite{Yao2021,Cherukara2018}, and modeling the continuous \num{3}D shape transition in heterogeneous samples \cite{Zhuang2022}. We train the VAE as described in \cite{Burgess2018}, using the code from \cite{Note11}.

UMAP is a dimensionality reduction technique based on manifold learning and topological data analysis and has been used with other spectrographic image data, such as Ronchigrams \cite{Li2019a} and Audio spectrograms \cite{Sainburg2020,Thomas2022}. We use UMAP with the default parameters and a fixed integer \emph{random\_state} for reproducibility.

The size of the low-dimensional representation for all mentioned methods was set to \num{1024}, identical to the dimensionality of the CLR-based representation space.

\bibliographystyle{naturemag}
\bibliography{library}

\begin{thebibliography}{10}
\bibitem{NoteDomainKnowledge}
\bibinfo{note}{Computer vision with diffraction images is the domain in our
  case}.
\bibitem{NoteTaskKnowledge}
\bibinfo{note}{A similarity analysis is the task in our case.}
\end{thebibliography}
% \\~\\
\subsection{Acknowledgments}
Excellent support has been provided by the ISG team of DPHYS at ETH Zürich. Funding is acknowledged from the SNF via Grant No. $200021$E\_$193642$ and the NCCR MUST. Further funding was provided by the Leibniz Society via Grant No. SAW/\num{2017}/MBI\num{4}, and from the DFG via Grant No. MO \num{719}/\num{14}-\num{2}.

\subsection{Author contributions}
Earlier versions of the contrastive-based diffraction imaging approach were tested by F.B. and D.G. under supervision of J.Z. and D.R.. Training and evaluation of all models was written by J.Z. and performed on the \emph{NUX-Noether} GPU high-performance-computer at ETH Zürich \cite{Note12}. B.L. and D.R. contributed to discussing the results. The manuscript was written by J.Z. with input from all authors.

\subsection{Competing interests}
The authors declare no competing financial interests.

\end{document}